\documentclass[aps,pra,showpacs,twoside,twocolumn,10pt]{revtex4-1}
\usepackage[colorlinks=true, citecolor=red, urlcolor=blue ]{hyperref}
\usepackage{times,epsfig,amssymb,amsfonts,amsmath,bm,subfigure,mathtools,amsthm,braket,soul,enumitem,color,physics,graphics,graphicx}
\usepackage[normalem]{ulem}

\begin{document}

\title{
%Sequential sharing of Quantum Network Nonlocality\\
%Identification of Network Nonlocality between Time-like Separated Observers\\
Limits of  network nonlocality probed by time-like separated observers}

\author{Pritam Halder$^{1}$, Ratul Banerjee$^{1}$, Shiladitya Mal$^{2, 3}$, Aditi Sen(De)$^{1}$}

\affiliation{$^1$ Harish-Chandra Research Institute, A CI of Homi Bhabha National
Institute,  Chhatnag Road, Jhunsi, Allahabad - 211019, India\\
		$^2$ Physics Division, National Center for Theoretical Sciences, Taipei 10617, Taiwan \\
$^3$ Department of Physics and Center for Quantum Frontiers of Research and Technology (QFort),
National Cheng Kung University, Tainan 701, Taiwan	
	}

\begin{abstract}
   
   In an entanglement swapping scenario, if  two sources sharing entangled states between three parties are independent, local correlations lead to a different kind of inequalities than the standard Bell inequalities, known as network local models. A highly demanding task is  to find out a way to involve many players nontrivially in a quantum network since measurements,  in general, disturb the system. To this end, we consider  here a novel way of sharing network  nonlocality  when two observers initially share  close to  a maximally entangled state.  We report that by employing unsharp measurements performed by one of the observers, six pairs  can  sequentially demonstrate the violation of bilocal correlations  while a maximum of two pairs of  observers can exhibit bi-nonlocality when both the observers perform unsharp measurements. We also find the critical noise involved in unsharp measurements in each round to illustrate the bi-nonlocality for a fixed shared entangled state as a resource.   We also establish a connection between entanglement content of the shared state, quantified via von-Neumann entropy of the local density matrix  for pure states and entanglement of formation for Werner states, and the maximum number of rounds showing violation of bilocal correlations. By reducing  entanglement content in the elements of the joint measurement by the third party, we observe that the maximum number reduces to two sequential sharing of bi-nonlocality even for the maximally entangled state when the settings at each side are  taken to be three and fixed.  
    
\end{abstract}

\maketitle

\section{Introduction}
\label{sec:intro}

Quantum resource states are shown to enhance capacities in transmitting both classical and quantum information  over classically known protocols which were later implemented  successfully between a single sender and receiver using different physical substrates  \cite{crypto1, crypto2, dc, tele@bennett, telerev15, tele-pan-photon}. However,  quantum technological developments  also require generalization and realization of these protocols in a multipartite domain involving several parties situated in distant locations, thereby building a quantum communication network \cite{Kimble08}. One of the prominent designs in this direction is the proposal of quantum repeaters, a combination of entanglement distillation and swapping,   \cite{Briegel98, zukowski93, Bose99, repeaterrev} by which entangled states are shared between observers separated by long-distance even in presence of noise.  A crucial step here is to  verify the resource content in the created  states. For shared entangled states, several identification schemes exist which include testing Bell inequalities \cite{Bell64, brunnerRMP14}, entanglement witnesses \cite{hhhh}, steering inequality \cite{guhnerev}. 

Apart from entanglement detection, studies of Bell's theorem  plays an important role 
in understanding quantum theory \cite{Bell64, brunnerRMP14}. It was shown that bipartite entangled pure states always violate some Bell inequalities \cite{Gisin91}. Over the years, Bell inequalities have also been generalized in  multipartite domains and hence become crucial to establish nonlocality in networks \cite{brunnerRMP14, Aditi05, Acin11,  Ratul20}. It has been realized that   if one considers that the sources which share entangled states are independent, a distinct kind of local realistic models can be constructed which are different from the standard Bell inequalities -- a violation of these inequalities confirms the nonlocality in networks \cite{branciard10, branciard12, branciard12a, tavakoli21, tavakolirev}. The simplest network is called the bilocal scenario  involving two independent sources which share two entangled states between three parties having three inputs and outputs --  a violation of the inequality that  confirms the impossibility of  local models  was introduced by Branciard-Rosset-Gisin-Pironio, referred to as BRGP inequality.   In these scenarios, several works have been carried out both in the chain of arbitrary length and in star networks for which different kinds of inequalities based on local models can be derived. Moreover, unlike the paradigms of standard Bell inequalities, independent resource consideration leads to  much more involved structures in the set of local correlations which include nonconvexity of the set. 

On the other hand,  projective  or sharp measurements can  reveal nonlocal correlations present in the states by collecting the statistics required for Bell inequalities although they can destroy the shared entanglement. On the contrary, weak  or unsharp measurements can serve both  purposes by providing a trade off relation between information gain and disturbance due to measurement \cite{Buschbook}. In recent times, the generalized measurements are shown to be important tools in various quantum information tasks like state discrimination \cite{peres88},  state tomography \cite{Derka98}, violation of Bell inequalities \cite{vertesi10}, randomness generation \cite{acin18}, detection of entanglement \cite{Guhne18}, creating multipartite entangled states \cite{pritam21}. At the same time, it was also found that a shared entangled state can be detected sequentially by  the violation of Bell and steering inequalities, device-dependent as well as -independent entanglement witnesses by a single observer or by both the observers where observers perform unsharp measurements \cite{Silva15, Mal16, Bera18, sasmal18, asmita19, Shenoy19, Colbeck20, Pawlowski20, Anwer2021, Srivastava21}.  Upto now, all the sequential scenarios considered assume that there is a single source which produces the shared state initially. 

In the present work, we  go beyond this picture (cf. \cite{sharingbinonnew, tavakolinew}). In particular, we consider  two independent sources which produce two noisy nonmaximally entangled states. In this situation, after joint measurement by the middle party, the other two observers' aim is to check nonlocal correlations and in the sharing scenario, the task is done by unsharp measurements (see Fig. \ref{fig:scheme}). In this paper, we consider two sequential scenarios 
-- (1) depending on  the unsharp measurement performed by one of the observers, they exhibit nonlocal correlations  sequentially by  obtaining violations of bilocal inequalities which we call  unidirectional sharing of bi-nonlocality; (2) both the observers find the critical unsharp parameters in each round to manifest nonlocal correlations  which we refer to as bidirectional sharing of bi-nonlocality. In both  scenarios, we establish a connection between the entanglement content of the shared state and the maximum number of cycles in which they are capable to demonstrate  bi-nonlocal correlations. We observe that in the unidirectional case, unlike standard Bell inequalities, the violation of bilocal inequalities can be observed with a maximum of \emph{six} rounds when the shared state is close to maximally entangled states, both for pure and noisy states. The maximum number reduces to two when both the observers wish to demonstrate the bi-nonlocality.   

 In an entanglement swapping protocol, the middle party has to carry out a joint entangling measurement on his/her parts. All the above results are obtained when Bell-basis measurements  are performed by the middle party.  In contrast, if the middle party performs a more general joint measurement, known as elegant joint measurement, another kind of inequality emerges to detect nonlocal correlations  having three input settings \cite{tavakoliPRL21}. We show that in this scenario, the maximum number  reduces to two  even for a unidirectional case with the noisy entangled state having high entanglement as an initial  resource while  we obtain that the sequential sharing is not possible when both the observers perform unsharp measurements, thereby reaching to a no-go theorem. Note that the results are true when each element in the elegant joint measurement basis contain a minimum amount of entanglement.

The paper is organized in the following way. In Sec. \ref{sec:prerequisites}, we introduce the bilocal  inequalities  both for Bell-basis and elegant joint measurements. We first present the recursion relation of different rounds involved in sequential sharing  ( SubSec. \ref{subsec:pro})  and two sequential scenarios, unidirectional (SubSec. \ref{subsec:uni} and bidirectional ones (SubSec. \ref{subsec:bi})  when the shared state is maximally  entangled pure states. We then consider the sequential scenario with noisy nonmaximally entangled states in Sec. \ref{sec:noisy}, thereby establishing a relation between entanglement
of the shared state and the maximal number of observers exhibiting network nonlocality. Going beyond Bell-basis measurement, and considering elegant join measurement, the sharing scenario changes drastically which will be discussed in Sec. \ref{sec:ejm}. We  finally conclude in Sec. \ref{sec:conclu}.

\section{Network Inequalities with different joint measurements}
\label{sec:prerequisites}

Let us briefly describe the network nonlocality, which is different from the standard Bell inequality. In an entanglement swapping scenario \cite{zukowski93, Bose99}, we assume that a single source creates two copies of a bipartite state, \(\rho\), which are shared between   Alice-Bob \((AB)\) and  Bob-Charu  (\(BC\)) pairs. After Bob's joint measurement on his parts, Alice and Charu can share an entangled state, \(\rho'\) whose entanglement content depends on the initial pairs and joint measurements by \(B\).  %\sout{\textcolor{magenta}{depending on the entanglement content of the initial pairs}}. 
Note that if the initial states are maximally entangled, the Bell-basis measurement at \(B\)'s node projects the  state between \(A\) and \(C\) into  maximally entangled. 
To detect entanglement between \(A\) and \(C\), several methods can be employed which include entanglement witness \cite{hhhh}, standard Bell inequalities \cite{Bell64, brunnerRMP14}, steering inequality to name a few. 

Instead of a single source,  we now assume that there are
%We  have a single source, in bilocal scheme, we have 
two independent sources \(S_1\) and \(S_2\), which emit two states characterized by hidden variables \(\lambda_1\) and  \(\lambda_2\) respectively and states corresponding to $\lambda_1$ $(\lambda_2)$ is shared by $AB$ $(BC)$ as shown in Fig. \ref{fig:scheme}. Here $A$ and $C$ have measurement settings labeled by $x$ and $z$ with outcomes $a$ and $c$, respectively while $B$ has a fixed measurement setting. In this situation, a new paradigm emerges known as bilocal scenario \cite{branciard10, branciard12, branciard12a, tavakoli21}.
In this paper,  we consider two kinds of measurements performed by $B$ --
%\begin{enumerate}
    %\item 
    1. \emph{Bell-basis measurement }(BSM), and
    %\item 
    2. \emph{elegant joint measurement }(EJM).
%\end{enumerate}
We will discuss about these bases in later part of this section. 
It has been established that for BSM, any bilocal model has to satisfy BRGP inequality~\cite{branciard12}  while different bilocal inequality is derived in case of EJM ~\cite{tavakoliPRL21}. We will now briefly discuss both of them.

\subsection{Local models based on BSM}

Let $A$ and $C$ have binary inputs and outputs, i.e., $x, z, a, c$ $\in$ $\{0,1\}$. However, $B$ has two bits of output, $\textbf{\emph{b}} = b^0 b^1 = 00, 01, 10, 11$ corresponding to the Bell-basis measurement,   $\{|\phi^{\pm}\rangle= \frac{1}{\sqrt{2}}(|00\rangle \pm |11\rangle), |\psi^{\pm}\rangle=\frac{1}{\sqrt{2}} (|01\rangle \pm |10\rangle\} $) and the  projectors for BSM are denoted by \(\Pi_{b^0 b^1}\). From the conditional probability, $P^{14}(a, b^0 b^1, c|x,z)$ obtained by three parties $A, B \text{ and } C$, after measurements being performed on their parts,  let us define the tripartite correlation, given by
\begin{eqnarray}
\label{eq:correlatorBRGP}
\langle A_x B^y C_z\rangle = \sum_{a, b^0 b^1, c} (-1)^{a+b^y+c}P^{14}(a,b^0 b^1,c|x,z),
\end{eqnarray}
where \(y\in \{0,1\}\).
Taking linear combinations of the above correlations,  we construct  two quantities, represented as
\begin{eqnarray}
\label{eq:BRGPtwo}
I^{14} &=& \frac{1}{4}\sum_{x,z} \langle A_x B^0 C_z\rangle,\\
J^{14} &=& \frac{1}{4}\sum_{x,z} (-1)^{x+z}\langle A_x B^1 C_z\rangle.
\end{eqnarray}
It was shown \cite{branciard10, branciard12} that
any bilocal model based on two independent sources would  satisfy the inequality, given by
%nonlinear Bell type inequality,
\begin{eqnarray}
\mathcal{B}:=\sqrt{|I^{14}|}+\sqrt{|J^{14}|} \leq 1. 
\label{eq:BRGP}
\end{eqnarray}
We will refer to the left hand side as BRGP function or expression. 

\subsection{Different bilocal scenario with EJM}

Instead of Bell-basis measurements by \(B\), we now consider a scenario in which \(B\) performs a joint entangling  measurement 
given by 
%EJM is constructed of 
%a family of equally entangled two qubit basis states 
$\{|\Psi_b^\theta\rangle\}_{b=1}^4$, parametrized by $\theta \in \{0,\frac{\pi}{2}\}$ with all elements of the basis being equally entangled \cite{tavakoliPRL21}. To construct the basis with this property, let us first write the pure states in cylindrical coordinates, representing the four vertices in a regular tetrahedron inside the Bloch sphere as
%subjected to the symmetry that four reduced states formed by tracing out any single qubit, produce a regular tetrahedron inside the Bloch sphere. 
%Four vertices of a regular tetrahedron formed inside the Bloch sphere is given by,
%\begin{eqnarray}
%\label{mb}
%\nonumber\vec{m_1} &=& (1,1,1),\hspace{2 cm}     \vec{m_2} = (1,-1,-1),\\
%\vec{m_3} &=& (-1,1,-1),\hspace{1.5 cm}     \vec{m_4} = (-1,-1,1).
%\end{eqnarray}
 %In cylindrical coordinates one can write the vertices as, $\vec{m_b}=\sqrt{3}(\sqrt{1-r_b^2}\cos\phi_b,\sqrt{1-r_b^2}\sin\phi_b,r_b)$, where $\vec{-m_b}$ denotes diametrically opposite direction in the Bloch sphere. We can write,
\begin{eqnarray}
\ket{\vec{\pm m_b}} = \sqrt{\frac{1\pm r_b}{2}}e^{-i\frac{\phi_b}{2}}\ket{0} \pm \sqrt{\frac{1\mp r_b}{2}}e^{i\frac{\phi_b}{2}}\ket{1}.
\end{eqnarray}
With the help of  them,  EJM basis reads as
\begin{eqnarray}
\label{ejmbasis}
\ket{\Psi_b^\theta}=\frac{\sqrt{3}+e^{i\theta}}{2\sqrt{2}}\ket{\vec{m}_b,-\vec{m}_b}+\frac{\sqrt{3}-e^{i\theta}}{2\sqrt{2}}\ket{-\vec{m}_b,\vec{m}_b}.
\end{eqnarray}
Notice that by varying $\theta$ from $0$ to $\pi/2$, one can reach from EJM to BSM (upto some local unitaries).

Unlike BSM,  $A$ and $C$ can choose to perform measurement out of three possible settings for each of them, i.e., $x,z \in \{0,1,2\}$, with binary outcomes $a,c \in \{0,1\}$. The four possible output of \(B\), representing the four vertices of the tetrahedron,   $\vec{m}_b$, given by %\begin{eqnarray}
%\label{mb}
%\nonumber
\(\vec{m_1} = (1,1,1)\), \(    \vec{m_2} = (1,-1,-1)\), 
\(\vec{m_3} = (-1,1,-1)\) and  \( \vec{m_4} = (-1,-1,1)\) can be labelled as 
%the four possible output of $B$ by 
the three-vector $\textbf{\emph{b}}=(b^1, b^2, b^3)$.  In joint measurement by \(B\), the bilocal inequality reads as
\begin{eqnarray}
\nonumber \mathcal{B_E}:=\frac{1}{3}\Bigg(\sum_{y=z}\langle B^y C_z\rangle &-& \sum_{x=y}\langle A_x B^y\rangle\Bigg)\\
&-& \sum_{x \neq y \neq z \neq x}\langle A_x B^y C_z \rangle \leq 3+5Z.\nonumber\\
\label{eq:EJM}
\end{eqnarray}
where $Z=\max\big\{{|\langle A_x \rangle|,|\langle A_x B^y \rangle|, \ldots ,|\langle A_x B^y C_z\rangle|}\big\}$ is the maximum of the absolute values of marginal and full correlators, which do not appear in $\mathcal{B_E}$.
As obtained in case of BSM in Eq. (\ref{eq:correlatorBRGP}),  we can also write the above  correlators in terms of the conditional probabilities emerged from experiments. For example, we have
\begin{eqnarray}
\label{eq:EJMcorr}
\langle A_x B^y C_z \rangle = \sum_{a,b^1,b^2,b^3,c} b^y(-1)^{a+c}  p(a,\textbf{\emph{b}},c|x,z).
\end{eqnarray}
The left hand side of (\ref{eq:EJM}) can be called Tavakoli-Gisin-Branciard (TGB) function. 
%Similarly, other correlators can also be written.

\begin{figure*}
\includegraphics[width=0.9\textwidth]{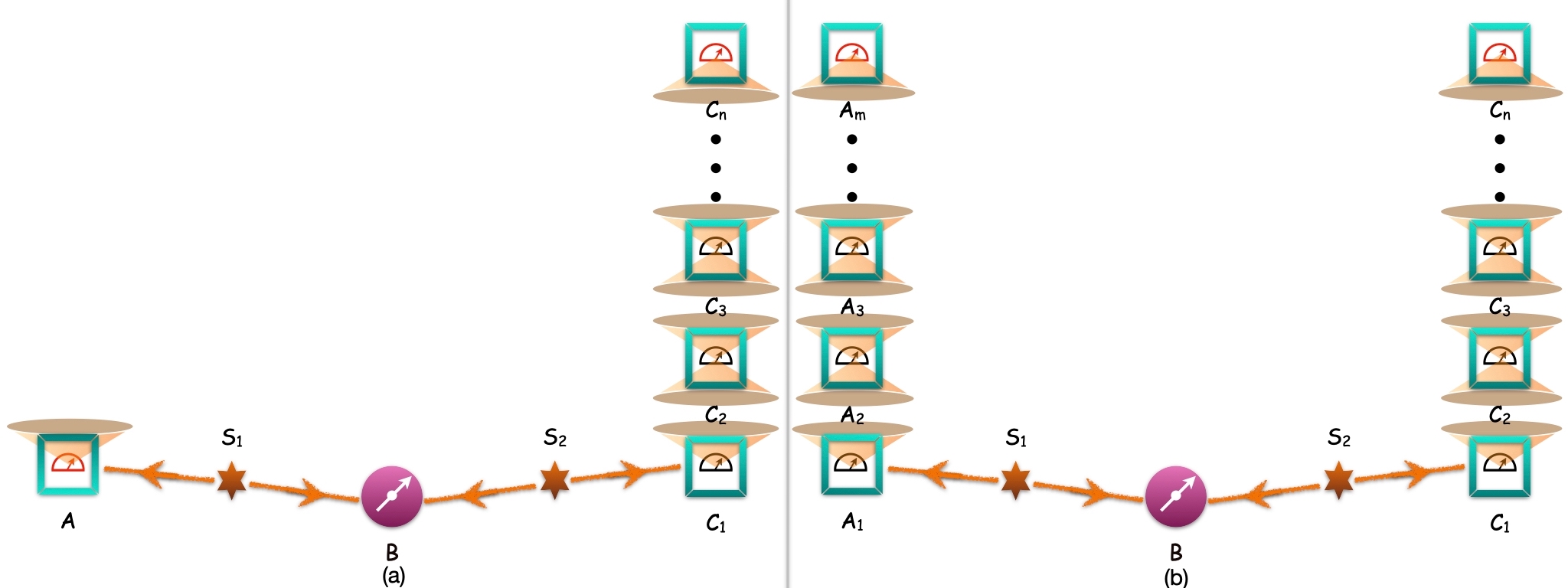}
\caption{(Color online.) Schematic diagram for sharing of quantum states sequentially. (a) Unidirectional where one of the observers performs weak measurement.  (b) Bidirectional sharing  of states in which both the observers perform weak measurements. In both the scenarios, nonlocal nature of states in each round is confirmed from the violation of network nonlocality (bilocal inequality).  }
\label{fig:scheme} 
\end{figure*}

\section{Sequential detection  of bi-nonlocality}
\label{sec:sequential}

Let us now set the framework of sharing bipartite quantum states sequentially. 
In this work, we consider following two   scenarios  (see Fig. \ref{fig:scheme} for schematics) -- \\
{\bf A. Unidirectional.} One of the parties performs unsharp measurements and  the other spatially separated observer does projective measurement  after \(B\)'s joint measurement in an entanglement swapping experiment.  \\
{\bf B. Bidirectional.}
Both the parties, i.e., \(A\)s and \(C\)s perform weak measurements, thereby disturbing the state minimally which occurs after the join measurement is completed at \(B\)'s end.

Before presenting the results, let us discuss the general protocol that will be followed to sequentially share and test quantum network nonlocality. 

%. Since we are interested to check nonlocality of quantum after arbitrary rounds, we consider that there are two sources, represented as \(S_i\) \((i=1, 2)\),   in which two arbitrary states \(\rho_i\) \((i=1, 2)\), are created and  one part of them are sent to \(B\) while the other parts are sent to \(A\) and \(C\)  from \(S_1\) and \(S_2\) respectively. If  \(B\) performs a projective entangling measurement on his two particles, \(A\) and \(C\) can share an entangled state depending on the entanglement of the original state, \(\rho\) -- a protocol known as entanglement swapping  \cite{zukowski93, Bose99}.  If the shared state is maximally entangled and \(B\) performs a Bell measurement, the resulting state between \(A\) and \(C\) can be shown to be maximally entangled. 

\subsection{General protocol for sharing bi-nonlocality in network}
\label{subsec:pro}

%Let consider the situation where  Bob (\(B\)) performs a 
After Bob's Bell-basis measurement, %the quantum correlations between Alice and Bob as well as Bob and Charu are destroyed since the measurement is projective. 
%with output $b_{0}b_{1}={00,01,10,11}$ respectively for $\phi^{\pm}=|00\rangle \pm |11\rangle,\psi^{\pm}=|01\rangle \pm |10\rangle $ .As performing a projective
 %will destroy the correlation between the shared state,
 Alices (Charus) denoted as $A^1,A^2,\ldots,A^m$ ($C^1,C^2,\ldots,C^n$)  perform weak measurements and send their parts of the qubits to the next Alice (Charu) which can capture the competition between information gain  and disturbance due to measurement. The measurement choices   of Alices (Charus) as inputs can be denoted as $x_1,x_2,\ldots, x_m $ ($z_1,z_2,\ldots, z_n$) and the corresponding measurement outcomes can be denoted as
$a_1,a_2,\ldots, a_m$ ($c_1,c_2,\ldots, c_n$). As mentioned before,  $\{x_m\},\{z_n\},\{a_m\},\{c_n\}\in \{0,1\}$. Suppose $A^m$ and $C^n$ choose their measurement directions with angle $\phi_m$ and $\theta_n$ respectively in the \(x-z\) plane, given by
\begin{eqnarray}
 A^{m}_{x_m} &=& \cos{\phi_{m}}\sigma_{z} - (-1)^{x_{m}}\sin{\phi_{m}}\sigma_{x}, \nonumber\\
 C^{n}_{z_n} &=& \cos{\theta_{n}}\sigma_{z} + (-1)^{z_{n}}\sin{\theta_{n}}\sigma_{x}. \end{eqnarray}
The projectors of Alices and Charus corresponding to their %measurement choice 
inputs and outputs are respectively  written as
\begin{eqnarray}
\label{eq:eigenpro}
\nonumber \Pi^{a_m}_{x_m} = \frac{(\mathbb{I}+ (-1)^{a_m} A^{a_m}_{x_m})}{2} \otimes \mathbb{I},
 \\
 \Pi^{c_n}_{z_n} =  \mathbb{I} \otimes \frac{(\mathbb{I} + (-1)^{c_n} C^{c_n}_{z_n})}{2}.  
 \end{eqnarray}
 Using these tools and concepts of unsharp measurements discussed in Appendix. \ref{sec:weak measurement}, we can put forward some simple steps to find the joint probability distribution $P(a_m,b^{0}b^{1},c_n|x_m,z_n)$, which will be required to check BRGP inequality (bilocal model with EJM) of the output states in each round. Let us enumerate each step of the protocol by Alices and Charus in details. 
 
 \begin{itemize}
     \item Initial state shared by Alice-Bob-Charu is denoted by $\rho = \rho_{A^{1}B}\otimes \rho_{BC^{1}}$. After Bob  performs the Bell-basis  measurement, the shared state between Alice and Charu depending on Bob's outcome can be written as
     \begin{eqnarray}
      \rho^{b^{0}b^{1}}_{A^{1}C^{1}}=\mbox{tr}_{B}[[\mathbb{I} \otimes \Pi_{b^{0}b^{1}}\otimes \mathbb{I}]\rho[\mathbb{I} \otimes \Pi_{b^{0}b^{1}}\otimes \mathbb{I}]^{\dagger}].
     \end{eqnarray}
     %\textcolor{red}{\(\rho_{b^{0}b^{1}}\) r mane ki?}
     
    \item Each Alice (till \((m-1)\) Alice, i.e., \(A^{m-1}\)) performs unsharp measurement according to the choice of the string $x_1,x_2,\ldots,x_{m-1}$ and the corresponding quality factors of the weak measurements, $F_1,F_2,\ldots,F_{m-1}$ \cite{Silva15}.  For a fixed round, say, \(k\), after \(A^k\)'s measurement, the part of the state is sent to the  next Alice, i.e., \(A^{k+1}\) without communicating the outcome. Hence the final transformed state between  $A^{m}$ and $C^{1}$ after \(m-1\) rounds of measurement by previous Alices, depending on all the previous Alice's measurement choices, can be written as
    \begin{eqnarray}
     &&\rho^{b^{0}b^{1}|x_1,x_2,\ldots,x_{m-1}}_{A^{m}C^{1}}=\\\nonumber &&\mathcal{W}_{x_{m-1}}(\mathcal{W}_{x_{m-2}}(\ldots\mathcal{W}_{x_{1}}(\rho^{b^{0}b^{1}}_{A^{1}C^{1}}) \ldots)),
    \end{eqnarray}
    where the map $\mathcal{W}_{x_{i}}$ is defined as
    \begin{eqnarray}
     &&\mathcal{W}_{x_{i}}(\rho) \\\nonumber 
     && = F_{i}\rho + (1-F_{i})\left(\Pi^{0}_{x_i}\rho (\Pi^{0 }_{x_i})^{\dagger}+ \Pi^{1}_{x_i}\rho(\Pi^{1}_{x_i})^{\dagger}\right).
    \end{eqnarray}
    
    \item In a similar fashion,  each Charu (till ($n-1$) Charu, i.e., \(C^{n-1}\)) performs unsharp measurement according to the choice of the string $z_1,z_2,\ldots,z_{n-1}$ and quality factors $F^{\prime}_1,F^{\prime}_2,\ldots,F^{\prime}_{m-1}$. 
    %and sends their part of the state to next charlie with out communicating the outcome.Now 
    The resulting state between $A^{m}$ and $C^{n}$ in this case reads as
    \begin{eqnarray}
     &&\rho^{b^{0}b^{1}|x_1,x_2,\ldots,x_{m-1},z_1,z_2,\ldots,z_{n-1}}_{A^{m}C^{n}}=\\\nonumber &&\mathcal{W}_{z_{n-1}}(\mathcal{W}_{z_{n-2}}(\ldots\mathcal{W}_{z_{1}}(\rho^{b^{0}b^{1}|x_1,x_2,\ldots,x_{m-1}}_{A^{m}C^{1}})\ldots )), 
    \end{eqnarray}
    where 
    %the map $\mathcal{W}_{z_{i}}$ is defined as
    \begin{eqnarray}
     &&\mathcal{W}_{z_{i}}(\rho) \\\nonumber 
      &&= F^{\prime}_{i}\rho + \left (1-F^{\prime}_{i})(\Pi^{0}_{z_i}\rho(\Pi^{0}_{z_i})^{\dagger}+ \Pi^{1}_{z_i}\rho (\Pi^{1}_{z_i})^{\dagger} \right).
    \end{eqnarray}
    
    \item In the last step,  $A^{m}$ and $C^{n}$ perform unsharp measurement according to the measurement choice $x_m$, and $ z_n$  with outcome $a_m$, and $c_n$ respectively. The post measurement state becomes
    \begin{eqnarray}
     &&\rho^{a_m,b^{0}b^{1},c_n|x_1,x_2,\ldots,x_{m},z_1,z_2,\ldots,z_{n}} \\\nonumber&& = \mathcal{W}^{c_n}_{z_{n}}(\mathcal{W}^{a_m}_{x_{m}}(\rho^{b^{0}b^{1}|x_1,x_2,\ldots,x_{m-1},z_1,z_2,\ldots,z_{n-1}}_{A^{m}C^{n}})).
    \end{eqnarray}
    where the corresponding operators,  $\mathcal{W}^{a_m}_{x_{m}}$ and $\mathcal{W}^{c_n}_{z_{n}}$ can be represented as
    \begin{eqnarray}
    && \mathcal{W}^{a_m}_{x_{m}}(\rho)\\\nonumber &&= \frac{F_{m}}{2}\rho +  \frac{(1+(-1)^{a_{m}} G_{m} - F_{m})}{2} (\Pi^{0}_{x_m}\rho (\Pi^{0}_{x_m})^{\dagger})  \\\nonumber
 &&+ \frac{(1-(-1)^{a_{m}} G_{m} - F_{m})}{2} (\Pi^{1}_{x_m}\rho (\Pi^{1}_{x_m})^{\dagger}).
    \end{eqnarray}

We are now ready to compute the joint probability distribution, given by
     \begin{eqnarray}
     &&P(a_{m},b^{0}b^{1},c_{n}|x_1,x_2,\ldots,x_{m},z_1,z_2,\ldots,z_{n}) \\\nonumber = &&tr(\rho^{a_m,b^{0}b^{1},c_n|x_1,x_2,\ldots,x_{m},z_1,z_2,\ldots,z_{n}}).
     \end{eqnarray}
     
     \item We also  require to consider the previous individual  probabilities by Alices and Charus measurement choices  and by performing average over all such measurement choices, we obtain
     \begin{eqnarray}
      &&P^{14}(a_{m},b^{0}b^{1},c_{n}|x_{m},z_{n})\\\nonumber
      &&=\sum_{x_1,..,x_{m-1},z_1,\ldots,z_{n-1}=0}^{1}P(x_1)\ldots P(x_{m-1}) P(z_1)\ldots\\\nonumber&&P(z_{n-1}) \times  P(a_{m},b^{0}b^{1},c_{n}|x_1,x_2,\ldots,x_{m},z_1,z_2,\ldots,z_{n})\\\nonumber
      &&= \sum_{x_1,\ldots,x_{m-1},z_1,\ldots,z_{n-1}=0}^{1}\frac{1}{2^{n+m-2}} \times \\\nonumber&& P(a_{m},b^{0}b^{1},c_{n}|x_1,x_2,\ldots,x_{m},z_1,z_2,\ldots,z_{n})
     \end{eqnarray}
    \item  After finding these joint correlations between $A^{m}, B$, and $C^{n}$, it is straightforward to find the conditions for which these correlations violate bilocal models using BRGP inequality. Specifically, the quantities required for BRGP inequality take the form as
    \begin{eqnarray}
     &&\langle A^{m}_{x_m} B^y C^{n}_{z_{n}}\rangle \\\nonumber
     &=& \sum_{a_{m}, b^0 b^1, c_{n}} (-1)^{a_{m}+b^y+c_{n}}P^{14}(a_{m},b^0 b^1,c_{n}|x_{m},z_{n}),\\\nonumber
     &&I_{m,n}^{14} = \frac{1}{4}\sum_{x_{m},z_{n}} \langle A^{m}_{x_m} B^y C^{n}_{z_{n}}\rangle,\\\nonumber
     &&J_{m,n}^{14} = \frac{1}{4}\sum_{x_{m},z_{n}}(-1)^{x_{m}+z_{n}} \langle A^{m}_{x_m} B^y C^{n}_{z_{n}}\rangle,\\\nonumber
    \end{eqnarray}
    and finally we obtain the condition on \((G, F)\)-pair such that 
%     Final BRGP condition for satisfying non-bi-locality becomes,
     \begin{eqnarray}
      \mathcal{B}(A^{m},B,C^{n}):=\sqrt{|I_{m,n}^{14}|}+\sqrt{|J_{m,n}^{14}|} > 1 .
     \end{eqnarray}
 \end{itemize}
 In case of unidirectional sharing, our aim is to find maximum \(m\) or \(n\) by performing projective measurement by the other party, i.e., by fixing the other index to be \(1\). On the other hand,  the maximum pair of \((m, n)\) will be found for the bidirectional situation.  Notice also that instead of Bell-basis measurement, if $B$ performs EJM, we can also compute the corresponding network inequality in each round by slightly modifying all the derivations obtained above.

 \subsection{Unidirectional sharing of bi-nonlocality}
\label{subsec:uni}

In the case of unidirectional sharing, our motivation is to find the criteria under which all temporally separated Charus share  a  bi-nonlocal correlation with a single Alice, thereby violating the BRGP inequality. Without loss of generality,  we can take $m=1$  and hence we assume that \(A^1\) does not perform weak measurement, i.e., she performs a sharp measurement with $G_{1}=1$. 

Let us first illustrate the situation when the shared states between \((A^1, B)\) and \((B, C^1)\) pairs  are maximally entangled and \(B\) performs the Bell-basis measurement.  After some manipulations, the general form of BRGP inequality between \(A^1, B\) and \(C^n\) can be written as
\begin{eqnarray}
\label{brgp1n}
  \mathcal{B}(A^{1},B,C^{n})&=& \sqrt{\left| I_{1,n}^{14} \right| }+\sqrt{ \left| J_{1,n}^{14} \right|}, 
   \end{eqnarray} 
   with 
   % \begin{widetext}
\begin{eqnarray} 
\label{eq:IJ1n}
 && I_{1,n}^{14} =  G^{\prime}_{n} \cos{\theta_{n}}\cos{\phi}\sum_{\{l_{i}\}=0}^{1}\prod_{i=1}^{n-1}(1+ (-1)^{l_{i}}F_{i}^{\prime})(\cos{2\theta_{i}})^{l_{i}}, \nonumber\\
 && J_{1,n}^{14} = \nonumber \\
 && G^{\prime}_{n}  \sin{\theta_{n}}\sin{\phi}\sum_{\{l_{i}\}=0}^{1}\prod_{i=1}^{n-1}(1+ (-1)^{l_{i}}F_{i}^{\prime})(\cos{2\theta_{i}})^{l_{i}}(-1)^{l_{i}} \nonumber\\
 \end{eqnarray} 
The expressions for a few rounds are mentioned in Appendix. \ref{sec:recur} which can clearly give us the idea to obtain optimal rounds for sharing.

\subsubsection{Optimal strategy to share bi-nonlocality}

After obtaining the BRGP expression in Eq. (\ref{eq:IJ1n}), for a fixed round, say, \(k\), we compute the minimum \(G'_k\) value  for which \(\mathcal{B}(A^1, B, C^k)\), a function of \(G'_k\), \(\theta_k\) and \(\phi\),   just starts violating the BRGP inequality. 
%We also note the corresponding \(\theta_k\)  and \(\phi\) values involved in projectors in Eq. (\ref{eq:eigenpro}). 
It can be easily confirmed by considering Eq. (\ref{eq:recurf}) that  minimum \(G'_k\) is obtained for any round, \(k\), when \(\theta_k = \phi = \pi/4\). For example,   \(\mathcal{B}(A^1, B, C^1)\)=1  leads to  condition for critical \(G^{\prime cr}_1\) as
\begin{equation}
 \sqrt{G^{\prime cr}_1} = \frac{1}{\sqrt{\left| \cos{\theta_{1}}\cos{\phi} \right| }+\sqrt{ \left| \sin{\theta_{1}}\sin{\phi}\right|} }
\end{equation}
Now it is obvious that $\min(\sqrt{G^{\prime cr}_1})= \frac{1}{\sqrt{2}}$ at $\theta_1 =\phi=\frac{\pi}{4}$. Putting these values of $G^{\prime cr}_1,\theta_1,\phi$ in
$\mathcal{B}(A^1, B, C^2)$ and demanding it to be unity, we can similarly show $\sqrt{G^{\prime cr}_2}$ is minimum with $\theta_2 = \frac{\pi}{4}$. Same arguments apply for all rounds and finally we can specify the optimal strategy of measurement at each round such that the state is minimally disturbed or probed to show bi-nonlocality with all $G^{\prime }_{i}>G^{\prime cr}_{i}$. Under this conditions, the  general form of \(\mathcal{B}(A^1, B, C^n)\) simplifies as
\begin{eqnarray}
 \mathcal{B}(A^{1},B,C^{n}) = 2 \sqrt{\frac{1}{2^{n}}\prod_{i=1}^{n-1}(1+F^{\prime}_{i})G^{\prime}_{n}}.
 \end{eqnarray}

\noindent\textbf{$\blacksquare$ Proposition I.} \emph{In the unidirectional sharing of bi-nonlocality,  when two independent sources produce two copies of  maximally entangled states, a single Alice can simultaneously violate BRGP inequality with a maximum of six  Charus provided Bell-basis measurement is performed by the middle party (Bob).
}\\
\textbf{Proof.} Let the maximum number of Charu showing bi-nonlocality with Alice be \(n\). To find the critical values of $G^{\prime cr}_i , \forall i=1,\ldots,n$, we need the solutions of $\mathcal{B}(A^1, B, C^n)=1, \forall \, i=1,\ldots,n$ which  lead to 
\begin{eqnarray}
&&G^{\prime cr}_1 = \frac{1}{2}, \\
\label{uni6}
&&G^{\prime cr}_{i+1}= \frac{2G^{\prime cr}_{i}}{1+F^{\prime cr}_{i}}. 
\end{eqnarray}
Simplifying this,  we get $ G^{\prime cr}_2 = 0.536, G^{\prime cr}_3 = 0.581, G^{\prime cr}_4 = 0.64, G^{\prime cr}_5 = 0.725, G^{\prime cr}_6 = 0.859, G^{\prime cr}_7= 1.135$. It immediately implies that only six Charus can satisfy Eq. (\ref{uni6}) for $0 < G^{\prime cr}_i \leq 1$ so that the resulting state  violates the BRGP inequality.
$\blacksquare$

In the succeeding sections, we will demonstrate that the BSM by Bob and entanglement content of the shared states are crucial to obtain the maximum number of rounds showing bi-nonlocality as depicted in Figs. \ref{fig:nme} and \ref{fig:werner}.  \\

\emph{Remark.} The optimal measurement settings, i.e., the values of \(\theta_k\ = \phi = \pi/4\) remain optimal also for the Werner state \cite{Werner'89} \(\forall k\) rounds (i.e., when the maximally entangled state is admixed with white noise).

\subsection{Bidirectional sharing of bi-nonlocality: Advantage in asymmetry}
\label{subsec:bi}

 As shown in the previous situation,  the optimal choice of measurement direction in this case also turns out  to be $\theta_{i}=\phi=\pi/4$. Using this,  we can generalise the BRGP violation  between $m$-th Alice and $n$-th Charu as
 \begin{eqnarray}
 &&\mathcal{B}(A^{m},B,C^{n}) = \\\nonumber 
 && 2 \sqrt{\frac{1}{2^{n+m-1}}\prod_{i=1}^{m-1}(1+F_{i})\prod_{i=1}^{n-1}(1+F^{\prime}_{i})G_{m}G^{\prime}_{n}}. 
 \end{eqnarray}
In other words,  $m$ number of Alices and $n$ number of Charus are said to be perfectly share bi-nonlocality bidirectionally if 
\begin{eqnarray}
 \mathcal{B}(A^{i},B,C^{j})>1 \,\,\,\forall i=1,\ldots, m; \, j = 1,\ldots,n. 
 \label{bi-cond}
\end{eqnarray}

\emph{Weak measurements with equal precision.} In this scenario, if we take the precision of the measurement at Alice and Charu's end to be equal, i.e.,  $G'_n = G_n$, we have the following results. \\   
\noindent\textbf{$\blacksquare$ Proposition II.} \emph{In the bidirectional sharing with equal precision in Alice and Charu's measurements in each round, a maximum number of Alice and Charu who can perfectly share bi-nonlocality sequentially  is two when the shared state is maximally entangled and the Bell-basis measurement is performed.
}\\
\textbf{Proof.} The proof is similar to Proposition I.  With the equality condition in  (\ref{bi-cond}), we get the criteria as
\begin{eqnarray}
&&G^{ cr}_1 = \frac{1}{\sqrt{2}}, \\
\label{bi2-2}
&&G^{ cr}_{i+1}= \frac{2G^{ cr}_{i}}{1+F^{cr}_{i}}. 
\end{eqnarray}
Calculating explicitly, we find $G^{ cr}_{2} = 0.828 , G^{ cr}_{3} = 1.06$. Therefore,  only two Alices can share perfect bi-nonlocality with two Charus having $0<G^{\prime cr}_i \leq 1$.
$\blacksquare$

\emph{Weak measurements with unequal precision.} Let us  now take precision of  unsharp measurements performed by Alices and Charus are unequal, i.e., \(G_m \neq G_n\). It is interesting  to check whether the situation is advantageous than the previous ones.

First, we find whether two Alices can share bi-nonlocality perfectly with more than two number of Charus. Without loss of generality, let us take sharpness parameter of second Alice to be $1$. Now taking equality sign in (\ref{bi-cond}), we obtain the following conditions, given by
\begin{eqnarray}
&&G^{ cr}_2 = 1, \\
&&G^{ \prime cr}_1 G^{ cr}_1= \frac{1}{2}, \\
\label{bi2-3}
&&G^{\prime cr}_{i+1}= \frac{2G^{\prime cr}_{i}}{1+F^{\prime cr}_{i}},\\
&&G^{cr}_{i+1}= \frac{2G^{cr}_{i}}{1+F^{cr}_{i}}.
\end{eqnarray}
Using these relation,  we immediately observe that $G^{\prime cr}_1,G^{\prime cr}_2,G^{\prime cr}_3 < 1$ and $G^{\prime cr}_4>1$. Thus in this asymmetric scenario,  \emph{two} Alices can show violation of BRGP inequality  with \emph{three} Charus. Similarly, for $m=3(4)$, we can show the maximum number of Charus can be $n=2(1)$.
Hence, we prove that at most $two$ Alices (Charus) can share  bi-nonlocality with a maximum of $three$ Charus (Alices) for a shared maximally entangled state and for BSM.
%which is similar with $m=3, n=2$ case.  

\section{Detecting bi-nonlocality sequentially with noisy nonmaximally entangled states }
\label{sec:noisy}

Instead of sharing a two-qubit maximally entangled states between Alice, Bob and Charu, let us consider the situation when  two independent sources can share noisy non-maximally entangled states  written as $\rho =\rho_{1}^{AB}\otimes \rho_{2}^{BC}$ with
$\rho_{1}^{AB} = v_{1}|\psi_{\alpha}\rangle\langle\psi_{\alpha}| + \frac{1-v_{1}}{4}{\mathbb{I}}$, 
and
$\rho_{2}^{BC} = v_{2}|\psi_{\beta}\rangle\langle\psi_{\beta}| + \frac{1-v_{2}}{4}{\mathbb{I}}$, having visibilities \(v_1\) ad \(v_2\) respectively and $|\psi_{\eta}\rangle = \sqrt{\eta}|00\rangle + \sqrt{1-\eta}|11\rangle$ with \(\eta = \alpha\) \text{or} \( \beta\). 

Following the similar prescription discussed in Sec. \ref{subsec:pro}, we can generalize the BRGP function between  Alice after \(m\) rounds and Charu after \(n\) rounds as
\begin{eqnarray}
 &&\mathcal{B}(A^{m},B,C^{n}) = \\\nonumber &&  \sqrt{\frac{1}{2^{n+m-1}}\prod_{i=1}^{m-1}(1+F_{i})\prod_{i=1}^{n-1}(1+F^{\prime}_{i})G_{m}G^{\prime}_{n}} \\\nonumber  &&\times \sqrt{v_{1}v_{2}} \times (1+2 \sqrt[4]{\alpha(1-\alpha)\beta(1-\beta)}).
 \label{eq:nonmaxnoisy}
\end{eqnarray}
Using the above recursion relation, we can find the minimum disturbance value at each round so that the shared state can show network nonlocality in maximum rounds.

\subsection{Bounds on sharing nonlocality between unidirectional time-like separated observers}

%\subsubsection{NME as resource}
{\bf NME as resource.} Let us first manifest the maximum number of rounds for which  bi-nonlocality can be shown when  both $AB$ and $BC$ share identical copies of  non-maximally entangled (NME) states, i.e., \(\rho_1^{AB}\) with \(v_1 =1\) (and similarly \(\rho_2^{BC}\) with \(v_2 =1\)) and \(\alpha = \beta\). 

We want to examine  the maximum number of sequential observers on Charu's ($C^m$) side can violate BRGP inequality with a single Alice ($A^1$) after \(B\)'s BSM. Using Eq. (\ref{eq:nonmaxnoisy}),  we can show that a maximum of six Charus can  sequentially demonstrate  bi-nonlocality with Alice when the initial resource is close to a ME state. 

\begin{figure}[ht]
\includegraphics[width=0.9\linewidth]{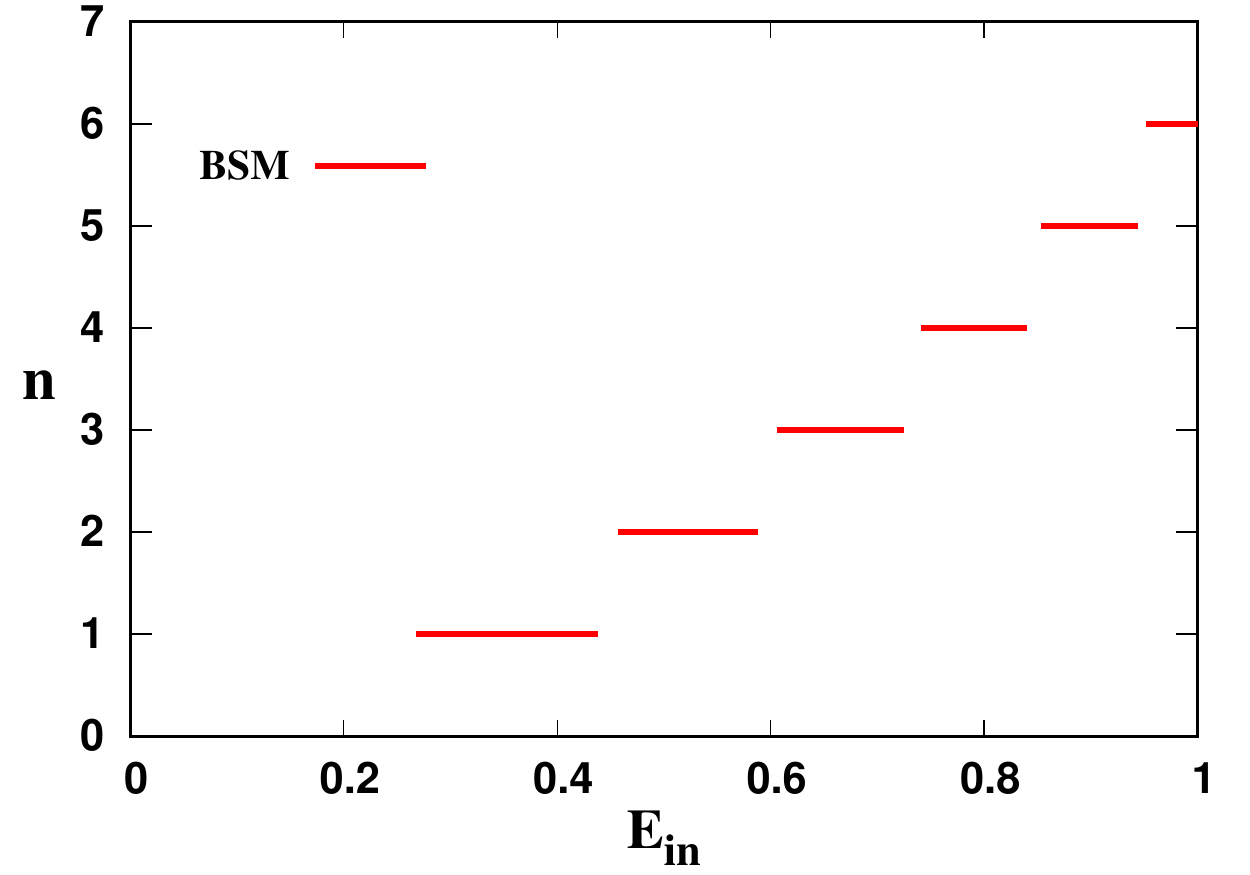}
\caption{(Color online.) \textbf{Maximum rounds vs. initial entanglement for NME states.}
The abscissa, $E_{in}$ denotes the entanglement of the initial resource states calculated in terms of von Neumann entropy of the local density matrices while the ordinate, $n$ signifies the number of Charu who can show violation of BRGP inequality with \(\theta_n = \phi = \pi/4\), thereby indicating the presence of nonlcoality in the shared pair.  The finite length of the steps implies that the maximum number of Charus who can demonstrate nonlocality with Alice after Bob's Bell-basis measurement remains fixed for a finite range of initial entanglement. Both the axes are dimensionless.}
\label{fig:nme} 
\end{figure}

The similar analysis also helps us to establish a connection between the entanglement content of the initial state, \(E_{in}\), quantified by the von-Neumann entropy of the local density matrices \cite{hhhh} and the maximum number of rounds. 
Specifically, we find that with the decrease of entanglement
%(Von Neumann entropy of $\rho_A$, i.e., $S(\rho_A)$ ) 
in the initial resource states, the number of  observers at one side (Charus) decreases as shown in Fig. \ref{fig:nme}. 
As shown in case of detection of entanglement sequentially via entanglement witness operators \cite{Bera18}, 
we also observe that along with the maximally entangled state, there is also other non-maximally entangled states, having entanglement \(E_{in} > 0.951\) which can  exhibit violation of BRGP inequality upto six rounds with \(\theta_i = \phi = \pi/4\) \(\forall i\). Also, the sharing of nonlocality is possible (i.e., the maximum of two Charus can demonstrate the violation of BRGP inequality) for $E_{in} > 0.456$. It is to be noted that the hierarchy  among NME states in this sharing scenario according to the violation of BRGP inequality is obtained with fixed settings, i.e., with \(\theta_i = \phi = \pi/4\) (\(i=1, \ldots, n\)) (cf. \cite{Gisin17}).
%two Charus can share bi-nonlocal advantage.
%show that maximum six Charlies can share the nonlocality for $S(\rho_A) > 0.95$. 
%Moreover, 
 % no Charu can show BRGP violation with Alice for $E_{in} < 0.267$ in our scenario.
  %, thereby showing violation of bilocal inequality as a weaker condition of detecting entanglement than other entanglement criteria like partial transposition, entanglement witnesses \cite{hhhh, tavakolirev}. \textcolor{red}{Bell inequality r moto somosto pure states violation dei na naki?} 

\textbf{Noisy entangled states as resource.}
In the unidirectional domain, similar analysis can also be carried out by taking two identical copies of noisy entangled states as initial resources with \(\alpha = \beta = 1/2\), i.e., the Werner states having \(v_1 = v_2\) \cite{Werner'89}.

\begin{figure}[ht]
\includegraphics[width=0.9\linewidth]{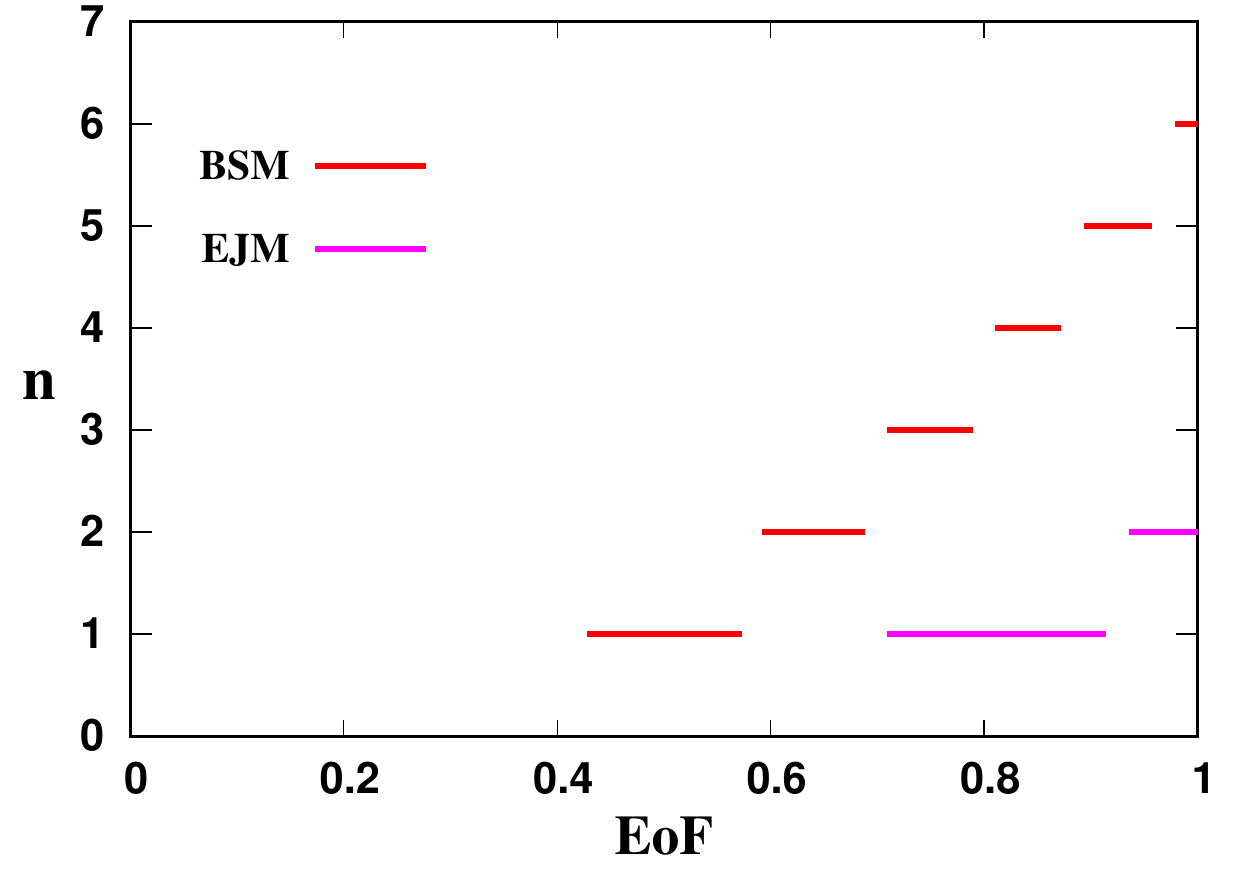}
\caption{(Color online.) \textbf{Maximum number of rounds against initial entanglement for Werner  states as initials.} 
The maximum number of rounds, \(n\)  by Charus (vertical axis) with respect to the initial entanglement quantified by entanglement of formation of the Werner state, \(EoF\) (horizontal axis). 
The implication is similar to Fig. \ref{fig:nme}. It indicates that  noisy entangled states can also behave as powerful as maximally entangled state in a sharing scenario. This observation can be important from the perspective of experiments where currently maximally entangled states can only be prepared with a certain but high visibility. Dark lines correspond to the scenario when Bob performs Bell-basis measurements while gray lines represent the elegant join measurements in Eq. (\ref{ejmbasis}) by Bob.    Both the axes are dimensionless.  }
\label{fig:werner} 
\end{figure}

Interestingly, we report that there exists a critical noise value upto which Charu can show bi-nonlocality with a single Alice in maximum six rounds 
(see  Fig. \ref{fig:werner}). In particular, if we calculate entanglement of formation \(EoF\) of the initial resource states \cite{Wootterseof}, we find that when $EoF > 0.978$, the maximum rounds that Alice-Charu-duo can sequentially share states which violate BRGP inequality is six. 
On the other hand, when $EoF < 0.428$,  Alice-Charu's state does not show violation even for a single round. 
%\textcolor{magenta}{ Note that the state is entangled when \(EoF >\)}.
The sequential protocol (i.e., minimum \(EoF\)  above which  two Charus can share bi-nonlocaliy with a single Alice) succeeds when \(EoF > 0.591 \). 

%\textcolor{red}{\(p=1/3\) mane \(EoF\) koto ektu dekhe bolo please. aar EoF kokhon theke n=2 pachho, ,seta-o dhukiye diyo.  }

\section{Sharing bi-nonlocality with EJM}
\label{sec:ejm}

Let us move to a scenario where Bob  performs EJM given in Eq. (\ref{ejmbasis}) and the corresponding bilocal inequality also modifies as in Eq. (\ref{eq:EJM}). Initially,  $A^1B$ and $BC^1$ share the Werner states, \(\rho_i=v_i\ket{\psi^-}\bra{\psi^-}+\frac{1- V_i}{4}\mathbb{I} \) with visibility $v_1$ and $v_2$ respectively.
Considering  measurement settings for obtaining the violation of bilocal models as $\{\sigma_x, \sigma_y, \sigma_z \}$ for Alice and Charus, the correlators  take the form as
\begin{eqnarray}
\nonumber && \langle A_x \rangle = \langle B_y \rangle = \langle C_z^n \rangle = \langle A_x C_z^n \rangle = 0,\\
\nonumber && \langle A_x B^y \rangle = -\frac{v_1}{2} \cos{\theta} \delta_{x,y},\\ \nonumber
&&\langle B^y C_z^n \rangle = \frac{v_2}{2} G_n^\prime \cos{\theta} \delta_{y,z} \prod_{i=1}^{n-1}K_i^\prime , \hspace{0.5cm} \text{where } K_i^\prime = \frac{1+2F_i^\prime}{3}\\ \nonumber
\langle && A_x B^y C_z^n \rangle \\ \nonumber
&&= -\frac{v_1 v_2}{2} G_n^\prime (1+ \sin{\theta}) \prod_{i=1}^{n-1} K_i^\prime \hspace{0.15cm} \text{if} \hspace{0.1cm} xyz \in \{123, 231, 312\} \\ \nonumber
&&= -\frac{v_1 v_2}{2} G_n^\prime (1 - \sin{\theta}) \prod_{i=1}^{n-1} K_i^\prime \hspace{0.15cm} \text{if} \hspace{0.1cm} xyz \in \{132, 321, 213\}\\ \nonumber
&& = 0 \hspace{4.5cm} \text{otherwise.}\\
\end{eqnarray}
Finally, the corresponding bilocal expression reads 
\begin{eqnarray}
\nonumber \mathcal{B_E}(A^1,B , C^n) = \frac{\cos{\theta}}{2}\bigg[v_1 &+&  v_2 G_n^\prime \prod_{i=1}^{n-1} K_i^\prime \bigg]\\ \nonumber &+& 3 v_1 v_2 G_n^\prime \prod_{i=1}^{n-1} K_i^\prime.\\
\end{eqnarray}
In the unidirectional case, when $v_1 = v_2 = v$, to show the violation of bilocal models at round $n$,  the weak measurement parameter has to satisfy
\begin{eqnarray}
G_n^\prime > \frac{6-v\cos{\theta}}{6v^2 + v\cos{\theta}} \prod_{i=1}^{n-1}\frac{1}{K_i^\prime}.  
\end{eqnarray}

If we consider $\theta=0$, one can show that the measurement settings of Alice and Charus considered above is optimal \cite{tavakoliPRL21}. 
%Here, we consider the EJM for which $\theta=0$ (all the basis elements of EJM have entanglement $E_{in}=0.354579$ ) only, where the measurement strategy of Alice and Charus are optimal in the direction given by $\sigma_x, \sigma_y, \sigma_z$.

\textbf{Maximally entangled state as resource}. When $v=1$, i.e., the resource state is maximally entangled,   at most two Charus can violate  (\ref{eq:EJM}) to share bi-nonlocality with a single Alice. Here, using $G_{1}^{\prime cr} = \frac{5}{7} \approx 0.714$, we find that $G_{2}^{\prime cr} \approx 0.893$. %In fact, they can share even if the state is NME, but for a very small-scale deviation from ME state, i.e., for $E_{in}>0.9988$.

\textbf{Werner state as resource}. %Taking Werner state $\rho_W = p\ket{\psi^-}\bra{\psi^-}+ \frac{1-p}{4}\mathbb{I}$ as resource,  
As shown in Fig. \ref{fig:werner}, when $EoF > 0.935$, two Charus can sequentially violate bilocal inequality involving EJM with a single Alice, while  even a single Charu cannot violate bilocal model for $EoF<0.689$.

If the shared state is NME, we find that the situation is much more involved. Taking \(\theta =0\) (i.e., each elements in the basis contains a very small amount of entanglement, \(E = 0.355\)), we observe that two Charus can violate the corresponding bilocal model with a single Alice sequentially only when $E_{in} > 0.998$.
Notice that  a single Alice-Charu duo cannot show bi-nonlocality when \(E_{in} < 0.976\).  
%non-bilocality even if the state is NME, but for $E_{in}>0.689$, i.e., a very small-scale deviation from ME state. For $E_{in}<9.976$ no Charu can have non-bilocal advantage.
Notice, moreover, that even  for a singlet states as initials, very less number of Charus can exhibit bi-nonlocality with \(A^1\) in comparison with BSM reported in the preceding section. It can be argued that such a disadvantageous situation emerges since each element of EJM$|_{\theta=0}$ contains a very low entanglement value, $E=0.355$ compared to the elements of BSM, having unit entanglement. 
%We believe varying $\theta$ and optimally choosing measurement strategies for Alice and Charus,  one may get larger number of Charus, who can satisfy .
It seems that to obtain the violation of $\mathcal{B_E}$, there is a competition between the entanglement content of the shared states and joint measurement basis and the choice of the  optimal measurement strategies by Alice-Charu pair.

\section{Discussion}
\label{sec:conclu}

In recent times, it has been established that unsharp measurements can provide certain benefits in quantum information processing tasks which cannot be reached by using projective (sharp) measurements due to its trade-off nature between the disturbance on the system and information obtained from the system. One prominent example is the sharing of entangled states in time-like separated observers which can be confirmed via the violation of Bell inequality, entanglement witnesses, steering inequality, etc. The violation of bilocal models of the resulting states after observers perform unsharp measurement sequentially are applied to detect nonlocality in the sharing scenario. 

Two kinds of sharing scenarios are considered -- unidirectional protocol where one of the observers 
performs unsharp measurement, and bidirectional process in which both the observers perform unsharp measurements. In the unidirectional scenario, we found that a maximum of six observers can exhibit bi-nonlocality when the shared state is maximally entangled. The maximum number of rounds for which the sharing of entangled states can be detected via the violation of bilocal models decreases with the decrease of entanglement content of the initial shared  states. We also observed that there exists a critical entanglement value of entanglement above which the multiple rounds of sharing bi-nonlocal states are possible. 
The situation changes drastically in the bidirectional case. In particular, the maximum number reduces to two when the shared state is close to maximally entangled states.  After completion of our work, we  notice that when both the observers share (noisy) maximally entangled states and want to employ network nonlocality by performing unsharp measurements in star and chain networks, a maximum of two rounds of detection for network nonlocal correlations is reported  \cite{tavakolinew} (cf. also \cite{sharingbinonnew}).

We also showed that the number of rounds where the sharing is possible also depends on the measurement performed by the middle party in an entanglement swapping experiment. Specifically, we found that instead of Bell-basis measurement, the elegant joint measurement \cite{tavakoliPRL21} has destructive effects on the protocol. It is possibly due to the fact that the elements of the elegant joint measurement except Bell-basis measurement contains a low amount of entanglement. 
We demonstrated that for a fixed elegant joint measurement,  the maximum of two rounds of sequential sharing is possible even in the unidirectional situation when the shared state is either maximally entangled  or close to the maximally entangled (in terms of non-maximally entangled pure state and maximally entangled state admixed with white noise).

\acknowledgements

We acknowledge the support from the Interdisciplinary Cyber Physical Systems (ICPS) program of the Department of Science and Technology (DST), India, Grant No.: DST/ICPS/QuST/Theme- 1/2019/23, and SM acknowledges the support from Ministry of Science and Technology, Taiwan (Grant No.MOST 110- 2124-M-002-012). PH and RB acknowledge the use of \href{https://github.com/titaschanda/QIClib}{QIClib} -- a modern C++ library for general purpose quantum information processing and quantum computing (\url{https://titaschanda.github.io/QIClib}), and the cluster computing facility at the Harish-Chandra Research Institute.  
\appendix

\section{Discussion: weak measurement}
\label{sec:weak measurement}
The notion of weak measurement can be captured by the formalism of unsharp measurement \cite{Mal16} in two outcome measurement scenario using the set of positive operator values measurement (POVM) or effective operators denoted as $E^{a}_{\lambda}= (\mathbb{I} + (-1)^{a} \lambda n^{i}\sigma^{i})/2 $ with $i=1,2,3$, $a\in{0,1}$ and $\lambda\in (0,1]$. Each POVM element can be written as a sharp projector mixed with white noise as
\begin{eqnarray}
 E^{a}_{\lambda,\overrightarrow{n}} = \lambda P^{a}_{\overrightarrow{n}} + \frac{1-\lambda}{2}\mathbb{I}, \\\nonumber
 P^{a}_{\overrightarrow{n}}=\frac{\mathbb{I} + (-1)^{a} n^{i}\sigma^{i}}{2}, \\\nonumber
 E^{0}_{\lambda,\overrightarrow{n}} + E^{1}_{\lambda,\overrightarrow{n}} = \mathbb{I}.
\end{eqnarray}
In this formalism, the outcome independent  unnormalized state of the system after measurement according to the Luder transformation rule, can be written as
\begin{eqnarray}
 \rho\prime & = &\mathcal{W}_{\overrightarrow{n}}(\rho) = \sqrt{E^{0}_{\lambda,\overrightarrow{n}}}\rho\sqrt{E^{0}_{\lambda,\overrightarrow{n}}} + \sqrt{E^{1}_{\lambda,\overrightarrow{n}}}\rho\sqrt{E^{1}_{\lambda,\overrightarrow{n}}} \nonumber\\
 &=& \sqrt{1- \lambda^{2}}\rho + \bigg(1- \sqrt{1-\lambda^{2}}\bigg)\big(P^{0}_{\overrightarrow{n}}\rho P^{0}_{\overrightarrow{n}} + P^{1}_{\overrightarrow{n}}\rho P^{1}_{\overrightarrow{n}}\big).\nonumber \\
\end{eqnarray}
Here, we identify $\lambda = G $ as the precision of the measurement and $\sqrt{1-\lambda^{2}} = F $ as the quality factor or the disturbance generated on the state due to the  performance of the measurement. The optimal  pointer condition for information gain-disturbance trade-off  is automatically satisfied by the unsharp formalism given by $F^2 + G^2 = 1$.

In the same way, we can get the outcome dependent unnormalized post measurement state as
\begin{eqnarray}
 &\rho\prime& = \mathcal{W}^{a}_{\overrightarrow{n}}(\rho) = \sqrt{E^{a}_{\lambda}}\rho\sqrt{E^{a}_{\lambda}}  \\\nonumber
 &=& \frac{F}{2}\rho +  \frac{(1+(-1)^{a} G - F)}{2} (P^{0}_{\overrightarrow{n}}\rho P^{0}_{\overrightarrow{n}}) \\\nonumber 
 &+&  \frac{(1-(-1)^{a} G - F)}{2} (P^{1}_{\overrightarrow{n}}\rho P^{1}_{\overrightarrow{n}}). 
\end{eqnarray}

\subsection{Recursion relation of BRGP function in unidirectional case}
\label{sec:recur}

 For simplicity, the general BRGP expression can be expressed for the rounds,  \(n = 1, 2, 3\) which finally leads to the recursion relation, given in Eq. (\ref{eq:IJ1n}) and the condition for sequential sharing. They are given by
\begin{widetext}
 \begin{eqnarray} 
 \label{eq:recurf}
&& \mathcal{B}(A^{1},B,C^{1})= \sqrt{G^{\prime}_{1}}\bigg\{\sqrt{\left| \cos{\theta_{1}}\cos{\phi} \right| }+\sqrt{ \left| \sin{\theta_{1}}\sin{\phi}\right|} \bigg\},  \nonumber \\
&& \mathcal{B}(A^{1},B,C^{2})= \sqrt{G^{\prime}_{2}}\bigg\{\sqrt{\left| \cos{\theta_{2}}\cos{\phi}((1+F_{1}^{\prime})+(1-F_{1}^{\prime})\cos{2\theta_{1}}) \right| }+\sqrt{ \left| \sin{\theta_{2}}\sin{\phi} ((1+F_{1}^{\prime})-(1-F_{1}^{\prime})\cos{2\theta_{1}})\right|} \bigg\},  \nonumber\\
&& \mathcal{B}(A^{1},B,C^{3}) = \nonumber\\
&& \sqrt{G^{\prime}_{3}}\bigg\{\sqrt{\left| \cos{\theta_{3}}\cos{\phi}((1+F_{1}^{\prime})(1+F_{2}^{\prime})+(1-F_{1}^{\prime})(1+F_{2}^{\prime})\cos{2\theta_{1}}+(1+F_{1}^{\prime})(1-F_{2}^{\prime})\cos{2\theta_{2}} +(1-F_{1}^{\prime})(1-F_{2}^{\prime})\cos{2\theta_{1}}\cos{2\theta_{2}}) \right| } \nonumber\\
&& +\sqrt{ \left| \sin{\theta_{3}}\sin{\phi}((1+F_{1}^{\prime})(1+F_{2}^{\prime})-(1-F_{1}^{\prime})(1+F_{2}^{\prime})\cos{2\theta_{1}}-(1+F_{1}^{\prime})(1-F_{2}^{\prime})\cos{2\theta_{2}}+(1-F_{1}^{\prime})(1-F_{2}^{\prime})\cos{2\theta_{1}}\cos{2\theta_{2}})\right|} \bigg\}.  \nonumber\\
\end{eqnarray}
\end{widetext}

\bibliography{ref}

\end{document}